\documentclass[12pt,preprint]{aastex}

\shorttitle{Formation of Supermassive Black Hole}
\shortauthors{Ebisuzaki et al.}

\begin{document}


\title{Missing Link Found? 
	--- The ``runaway'' path to supermassive black holes}


\author{
	Toshikazu Ebisuzaki\altaffilmark{1},
	Junichiro Makino\altaffilmark{2},
	Takeshi Go Tsuru\altaffilmark{3},
	Yoko Funato\altaffilmark{4},
	Simon Portegies Zwart\altaffilmark{5},
	Piet Hut\altaffilmark{6},
	Steve McMillan\altaffilmark{7},
	Satoki Matsushita\altaffilmark{8},
	Hironori Matsumoto\altaffilmark{9, 10},
	Ryohei Kawabe\altaffilmark{11}}

\altaffiltext{1}{
	Advanced Computer Center, RIKEN , 2-1 Hirosawa, Wako 351-0198, Japan}

\altaffiltext{2}{
	Department of Astronomy, University of Tokyo, 7-3-1 Hongo,
	Bunkyo-ku,Tokyo 113-0033, Japan}

\altaffiltext{3}{
	Department of Physics, Faculty of Science, Kyoto University,
	Kitashirakawa, Sakyo, Kyoto 606-8502, Japan}

\altaffiltext{4}{
	General Systems Sciences, Graduate Division of International and
	Interdisciplinary Studies, University of Tokyo, 3-8-1 Komaba, 
	Meguro-ku, Tokyo 153,8902, Japan}

\altaffiltext{5}{
	Massachusetts Institute of Technology, Cambridge, MA 02139, USA}

\altaffiltext{6}{
	Institute for Advanced Study, Princeton, NJ 08540, USA}

\altaffiltext{7}{
	Department of Physics, Drexel University, Philadelphia, PA 19104, USA}

\altaffiltext{8}{
	Harvard-Smithsonian Center for Astrophysics, P.O. Box 824, Hilo,
	HI 96721-0824}

\altaffiltext{9}{
	Department of Earth and Space Science, Osaka University, 1-1
	Machikaneyama, Toyonaka, Osaka 560-0043, Japan}

\altaffiltext{10}{
	Center for Space Research, Massachusetts Institute of
	Technology, 77 Massachusetts Avenue, MA 02139-4307, USA}

\altaffiltext{11}{
	National Astronomical Observatory, Ohsawa, Mitaka, 181-8588, Japan}


\begin{abstract}
Observations of stellar kinematics, gas dynamics and masers
around galactic nuclei have now firmly established that many galaxies
host central supermassive black holes (SMBHs) with masses in the range
$10^6 \sim 10^9$M$_{\odot}$. However, how these SMBHs formed is not
well understood. One reason for this situation is the lack of
observations of intermediate-mass BHs (IMBHs), which could bridge the
gap between stellar-mass BHs and SMBHs. Recently, this missing link
(i.e., an IMBH) has been found in observations made by the ASCA and
the  Chandra of the central region of the starburst galaxy M82
\citep{MT99, PG99, MT01, Ka01}. Subsequent 
observations by SUBARU have revealed that this
IMBH apparently coincides with a young compact star cluster.  Based on
these findings, we suggest a new formation scenario for SMBHs.  In
this scenario, IMBHs first form in young compact star clusters through
runaway merging of massive stars. While these IMBHs are forming, the
host star clusters sink toward the galactic nucleus through dynamical
friction, and upon evaporation deposit their IMBHs near the galactic
center. The IMBHs then form binaries and eventually merge via
gravitational radiation, forming an SMBH.
\end{abstract}




\section{Introduction}

There is rapidly growing evidence for SMBHs in the centers of many
galaxies (for a review see \citet{KR95}). There
are too many examples to list here; indeed, there are only a few
galaxies for which observations indicate that a central SMBH does not
exist\citep{KM93}.  Many authors have pointed out that the mass of the
central black hole $m_{BH}$ correlates linearly with the mass of bulge 
$M_b$, {\it
i.e.}, the ratio of $m_{BH}$ to $M_b$ is almost constant (0.002
\citep{KR95} to 0.006\citep{Mg98}). This suggests that the formation of
the central BH is somehow related to the formation of the bulge.

The formation mechanism of SMBHs is not well understood. Our
theoretical understanding has not advanced much beyond the scenarios
described by Rees\citep{Rs78,Rs84} in the early 1980s.  In the famous
diagram by Rees, there were basically two paths from gas clouds to
massive black holes. The first is direct monolithic collapse, the
second is via the formation of a star cluster, with subsequent runaway
collisions leading to BH formation. Previous numerical studies,
however, have demonstrated that neither path is likely. In the first,
a massive gas cloud is more likely to fragment into many small clumps,
in which stars then form, so direct formation of a massive BH from a
gas cloud seems unlikely.  In the second, stellar dynamics in star
clusters does not easily lead to the formation of massive black holes. 
A number of low-mass black holes (masses around 10M$_{\odot}$) are
formed via the evolution of massive stars, and these black holes do
indeed sink to the center of the cluster through dynamical friction
and form binaries by three-body encounters.  However, recent $N$-body
simulations\citep{PZM00} have demonstrated that practically all of
these black-hole binaries are ejected from the cluster by recoil
following interactions with other black holes (or BH binaries) before
they can merge through gravitational radiation.

\section{IMBHs in M82}

\citet{MT01} have identified nine bright compact X-ray
sources in the central region of M82 using recent Chandra data. The
brightest source (No.\ 7 in their Table 1) had a luminosity of $9
\times 10^{40} {\rm erg s}^{-1}$ in Jan 2000, corresponding to a black
hole with a minimum mass of $700 {\rm M}_{\odot}$ (assuming emission
at the Eddington limit). It probably consists of a single compact
object, as its X-ray flux shows rapid time variation\citep{MT01}.  This
is the first detection of a BH with a mass much greater than $100 {\rm
M}_{\odot}$ but much less than $10^6 {\rm M}_{\odot}$. Among the
eight other sources, at least three (5, 8 and 9) have Eddington masses
greater than $30 {\rm M}_{\odot}$.

\citet{Mts00} observed the same region in using
Nobeyama Millimeter Array and found a huge expanding shell 
of the molecular gas. They estimated the age and kinetic energy of the 
shell to be around 10 Myr and $10^{55}$ erg, which suggest a strong
starburst took place a few Myrs ago. 

\citet{Ha01} observed the same region in the IR
(J, H, and K$^\prime$-band)  using the CISCO
instrument on the SUBARU telescope\citep{Ha01}. They identified a
number of young compact star clusters, at least four of them
coinciding with the X-ray sources within the position uncertainty of
Chandra and SUBARU. The SUBARU field of view is smaller than that of
Chandra, and three of the X-ray sources are outside of the IR
field. Thus, four out of six bright X-ray sources in the IR field of
view have been identified with bright star clusters.  The logical
conclusion from these observations is that most of Chandra X-ray sources,
including the brightest one with an Eddington mass of $700 {\rm
M}_{\odot}$, are formed in star clusters.

Therefore, we now have two important observational results. The first
is that a BH with intermediate mass ($100< M_{bh}/M_{\odot}<10^6$) has
been found. The second is that it coincides with a young compact star
cluster. In the following, we discuss how these findings change
our understanding of the formation of supermassive BHs. We first
discuss how IMBHs can be formed in young compact star clusters, then
how IMBHs might grow into SMBHs.

\section{IMBH formation through runaway growth}

Here we consider some formation mechanisms for IMBHs in star clusters
(see figure 1). An obvious way to form an IMBH is the collapse of a
supermassive star. However, this possibility seems to have little
observational or theoretical support.  Observationally, no stars with
masses larger than $\sim200 {\rm M}_{\odot}$ are known. The most
luminous star currently known is the Pistol Star\citep{Fg98}, with an
estimated present mass of less than $200 {\rm M}_{\odot}$. Recent
radio observations of this star have detected signs of a strong
stellar wind\citep{Langetal1999}, which is consistent with the
theoretical prediction that such massive stars are dynamically
unstable\citep{Fg98}. We therefore regard it as unlikely that an IMBH
with mass exceeding $700 {\rm M}_{\odot}$ could have formed by the
collapse of a single massive star.

An alternative possibility is the formation and growth of massive
stars (and IMBHs) through successive merging. Massive stars in star
clusters have higher merger rates than less massive cluster members
(or field stars), for the following three reasons. First, they have
larger geometrical cross sections, and these are further enhanced by
gravitational focusing. Second, because of mass segregation, these
stars tend to be found in the high-density region near the cluster
center. Third, the most massive stars at the center tend to form
binaries through three-body interactions, and binary membership
greatly increases their chances of merging. A significant fraction of
binary-single star encounters lead to complex resonances during which
very close encounters can occur. Many such encounters lead to physical
collisions between stars \citep{HutInagaki1985,McMillan1986}. If these
effects are strong enough, we expect that a  ``merging
instability''\citep{Lee1987}, or a runaway growth of the most massive
star,  will occur in the cluster core.

\citet{Lee1987} performed Fokker-Planck simulations of globular
clusters with merging of stars taken into account. He included 
the first two of the three effects mentioned above, and concluded 
that the merger instability is
unlikely unless the mass of the globular cluster is very large
($N>10^7$, where $N$ is the number of stars). Recently, however,
$N$-body simulations carried out by \citet{PZ99}
have demonstrated that runaway merging can take place in much smaller
systems containing only $\sim 12,000$ stars. The primary reason for
this discrepancy is that, in $N$-body simulations, all three of the
effects described above are automatically included, whereas in the
Fokker-Plank calculation by Lee the third was not modeled. A second
reason is that Portegies Zwart et al. considered a young, compact
cluster, similar to R136 in the LMC or the Arches and Quintuplet
systems in the galactic center, while Lee considered normal globular
clusters. Thus, the dynamical timescale was much shorter in the models
considered by Portegies Zwart et al., allowing merging to occur before
stellar evolution eliminated the most massive stars.

Portegies Zwart et al. found that, in one case, the most massive star
experienced more than ten collisions and reached a mass of around $200
{\rm M}_{\odot}$ before experiencing a supernova. There is
considerable uncertainty as to how much mass would remain as a BH
after the supernova explosion of such a massive star, but it is quite
likely that the remnant black hole would still be one of the most
massive objects in the cluster, and that the runaway merging process
would continue. Although the geometrical cross section of a BH is
small, ``merging'' would take place when a star approached within its
tidal radius, leading to a relatively large merger cross section.

In order for runaway merging to occur, the dynamical friction
timescale for the most massive stars must be short enough that they
can sink to the center during their lifetime of several Myr.  The
dynamical friction timescale may be expressed as\citet{BT87}:
\begin{equation}
t_{fric} = \frac{1.17}{\log \Lambda}\frac{r^2v_c}{Gm} \simeq 2.7
\times 10^9 \left(\frac{r}{10 {\rm pc}}\right)^2
\left(\frac{r_h}{10 {\rm pc}}\right)^{-1/2}
\left(\frac{M}{10^6 M_{\odot}}\right)^{1/2}
\left(\frac{20 M_{\odot}}{m}\right) {\rm yr},
\label{eq:df}
\end{equation}
where $\log \Lambda$ is the Coulomb logarithm, $G$ is the
gravitational constant, $v_c$ is the local velocity dispersion, $r$ is
the distance from the center of the cluster, $r_h$ and $M$ are the
half-mass radius and the total mass of the cluster, and $m$ is
the mass of the star.

In the following, we consider how the dynamical friction works in the
cluster found in 
M82. From the infrared luminosity, \citet{Ha01}
estimate that the total mass of the cluster is $\sim 5 \times 10^6{\rm
M}_{\odot}$.  They also estimated the seeing-corrected radius of the
cluster as 5 pc. The dynamical friction  timescale is
around 800 Myr for stars at the half mass radius of the cluster. This looks
too
long, but in fact it is not. As can be seen from equation
(\ref{eq:df}), the timescale depends rather strongly on $r$.  For
$r<r_h/10$, a volume which still contains about 5\% of the total
cluster mass, the dynamical friction timescale is less than 10 Myr. We
therefore conclude that there is time for  a fair fraction of the most
massive stars in the cluster to sink to the cluster center and
undergo runaway merging before exploding as supernovae.

After the BH has become much more massive than other cluster members,
it forms a cusp near the cluster center \citep{BahcallWolf1976}, and
continues to consume other stars. Unfortunately, no realistic
simulations of this phase of the evolution are available. 
\citet{MarchantShapiro1980} performed Monte-Carlo simulations of
this stage for a simplified cluster containing $3\times 10^5$
solar-mass stars and one 50 solar-mass seed BH. They found that the BH
mass jumped to over $10^3 M_{\odot}$ (0.3\% of the cluster mass)
almost immediately after they put the BH into the system. After this
initial rapid growth, a slower phase ensued, with a doubling timescale
comparable to the relaxation time of the cluster.  Their result should
be regarded as a lower limit on the BH growth rate, since realistic
effects, in particular the presence of a mass spectrum, would greatly
enhance the accretion rate.  Taking these effects into account, it
seems safe (even conservative) to suppose that 0.1\% of the total
cluster mass accretes to form a $\sim 5000M_{\odot}$ central BH in a
few Myr.

As stated above, there are more than 10 bright star clusters in the
vicinity of the IMBH host cluster in M82, some of them apparently
hosting small BHs. Their age is around 10 Myr\citet{Ha01}.  Also, the
starburst in
M82 is a long-duration event, having started at least 200 Myr
ago\citet{deG2001}. If
we assume that the clusters are formed in a constant rate, we would
conclude that around 200 clusters have been formed. The star formation 
rate certainly has not been constant, so there is considerable
uncertainty in our estimate. Even so, we believe it is 
safe to assume that around one   hundred clusters similar to our host
cluster have formed in total, and that a considerable fraction of them
host IMBHs.

\begin{figure}[htbp]
\begin{center}




\includegraphics[width=80mm]{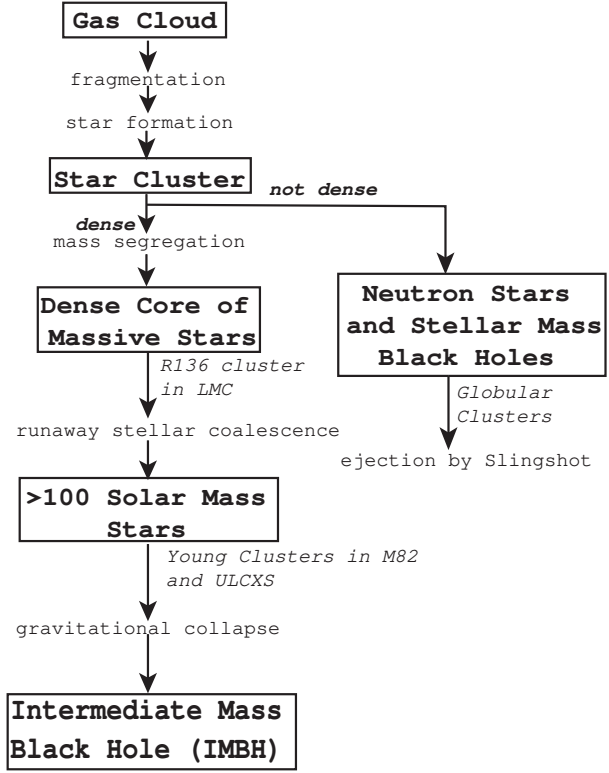}

\caption{Schematic diagram of the formation process of an intermediate-mass
black hole (IMBH). A gas cloud fragments to form many less
massive clouds as it cools by radiation. Many stars are formed through
this fragmentation, and a star cluster comes into being. There are two
possible evolutionary paths for this cluster, depending on its stellar
density.  If the star cluster is so dense that stellar mass
segregation is faster than stellar evolution for the most massive
stars (time scale $\sim 10^6$ yr), those stars sink to the cluster
core by dynamical friction, and form a dense inner core of massive
stars at the cluster center. In this inner core, the massive stars
undergo runaway stellar merging and a very massive star with mass
exceeding 100 $M_{\odot}$ forms. This massive star eventually collapse
into a black hole, which continues to grow by swallowing
nearby massive stars.  If the cluster is not dense enough for mass
segregation to occur in 10 Myr, massive stars evolve into compact
stellar remnants such as neutron stars and stellar-mass black holes
($\sim 10 M_{\odot}$). Those stellar remnants slowly sink to the
cluster center, since they are heavier than other stars in the system,
and eventually form binaries. Successive three-body interactions made
these binaries more tightly bound, and eventually they are ejected
 from the cluster by slingshot mechanism. 
}
\label{fig:IMBH}
\end{center}
\end{figure}

\section{Building up the central SMBH}

The growth rate of the IMBH in a star cluster slows once massive stars
disappear (after $\sim 100$ Myr). What happens to the cluster and the
IMBH after the initial rapid growth phase? In the discussion below, we
assume a cluster mass of $5\times 10^6$ M$_{\odot}$ and a half-mass radius
of
5pc.

The cluster is subject to two evolutionary processes: evaporation
through two-body relaxation and orbital decay (sinking) via dynamical
friction.  Evaporation is driven partly by thermal relaxation and
partly by stellar mass loss. 
\citet{PortegiesZwart2000} estimated that the evaporation timescale
for a Roche-lobe-filling compact star cluster is around 2-3 half-mass
relaxation times, which is of the order of a few Gyr for our star
clusters.  Rewriting equation (\ref{eq:df}) using appropriate scaling
for this case, we find that timescale on which the cluster sinks to
the galactic center via dynamical friction is
\begin{equation}
t_{fric} \simeq 6 \times 10^8
\left(\frac{r}{1 {\rm kpc}}\right)^2
\left(\frac{v_c}{100 {\rm km s^{-1}}}\right)
\left(\frac{5 \times 10^6 M_{\odot}}{m}\right).
\label{eq:df2}
\end{equation}
Clusters initially within 1 kpc of the galactic center can therefore
reach the center within one Gyr. Note that an IMBH can reach
the galactic center only if its host cluster can sink to the center
before it evaporates. If the cluster dissolves before significant
orbital decay occurs, the timescale for the IMBH to fall to the center
increases greatly.

According to our estimate in the previous section, around 100
compact clusters have formed close to the center of M82 in the last
200 Myrs. 
If we assume that half of these clusters contain $5000 {\rm
M}_{\odot}$ IMBHs, and that these IMBHs actually merge, then the total
BH mass at the center of the galaxy would be at least $2.5 \times 10^5{\rm
M}_{\odot}$, high enough to be called a supermassive BH,
if we take into account the relatively small total mass of M82. 

Having demonstrated that $5000 {\rm M}_{\odot}$ IMBHs can form and
reach the galactic center in a reasonable timescale, we now turn to the
question of whether the multiple IMBHs at the center can
merge.  \citet{Begelmanetal1980} discussed the evolution of a supermassive
BH binary at the center of a galaxy, taking into account dynamical
friction from field stars and energy loss via gravitational
radiation. They found that the merging timescale depends strongly on
mass, and for the very massive BHs in which they were interested,
merging took much longer than a Hubble time.

There are two reasons why they obtained such a long timescale. The
first is that they considered SMBHs with masses exceeding $10^8
M_{\odot}$, and the timescale for orbital decay via gravitational
radiation is proportional to the mass, for given orbital velocity. The
second reason is that they assumed that ``loss-cone depletion'' would
occur. In other words, they assumed that stars approaching close
enough to the BH binary to interact were supplied only though two-body
relaxation, which is a very slow process in large elliptical
galaxies. 

However, for the IMBHs we consider here, the timescale for merging
through gravitational radiation is many orders of magnitude shorter
than that for the SMBHs considered by Begelman et al. One reason is
simply that the IMBH mass is many orders of magnitude smaller.  An
IMBH of mass $10^4 M_{\odot}$ has a timescale $10^4$ times shorter
than a $10^8 M_{\odot}$ SMBH when compared at the same orbital
velocity.  A second reason is that loss-cone depletion is not as
effective as originally assumed, at least for relatively small BH
mass. Therefore IMBHs can reach high orbital velocities in short
timescales.  Recent extensive numerical simulations
\citep{Mk93,Mk97,QM97} have shown that the hardening of the BH binary
through dynamical friction is in fact several orders of magnitude
faster than the prediction from loss-cone arguments. A third reason
for the long SMBH merger time is the fact that Begelman et al. ignored
the effects of eccentricity, which can significantly reduce the
merging time scale.  Thus, we can safely conclude that the merging
time scale for IMBHs with masses less than $10^4 {\rm M}_{\odot}$ is
no more than a few Myr.

Since we expect many IMBHs to fall to the center of the galaxy on a
timescale of 100 Myr, in principle, slingshot ejection might remove
some of them from the region. However, once one BH has become more
massive than typical infalling BHs, it becomes extremely unlikely that
it will be ejected, since the kick velocity is inversely proportional
to the mass (because of momentum conservation). Thus, even though some
of the infalling BHs might be ejected, the central BH would continue
to grow. Figure 2 summarizes the discussion in this section.

\begin{figure}
\begin{center}

\includegraphics[width=80mm]{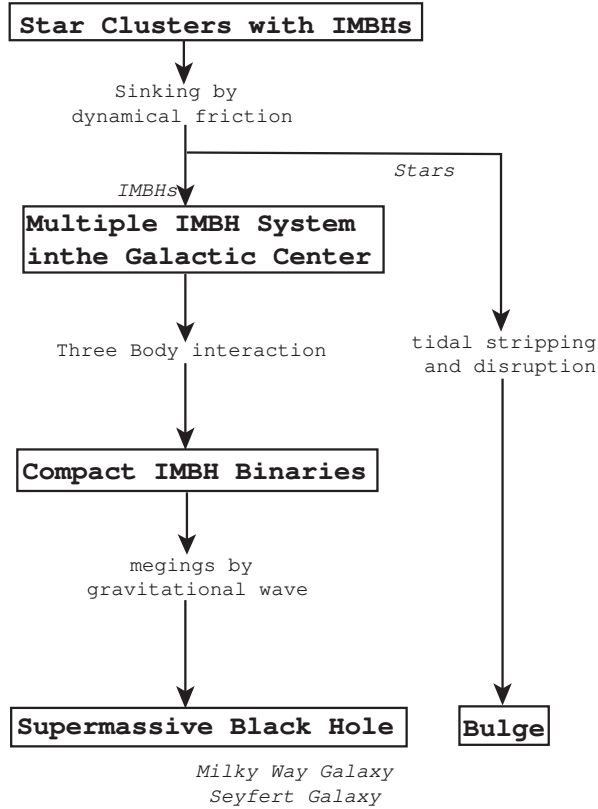}

\caption{Schematic diagram of the formation of supermassive black
holes from star clusters containing intermediate-mass black holes (IMBHs).
The star clusters sink to the galactic center by dynamical friction.
The tidal field of the parent galaxy strips stars from the outskirts
of the cluster. Those stripped stars ultimately become the galactic
bulge. The IMBHs carried to the center by the star clusters form a
multiple IMBH system at the center of the galaxy. IMBH binaries are
formed and become harder and harder by three-body interactions with
other IMBHs. Eventually, they merge into one or more massive black
holes through gravitational radiation. Successive mergings of IMBHs
form a supermassive black hole with a mass of $\sim10^6$M$_{\odot}$}
\label{fig:SMBH}
\end{center}
\end{figure}

\section{Discussion}

In this paper, we have discussed the implications for our
understanding of the SMBH formation mechanism of the recent discovery
of an IMBH in M82. Our conclusion is that the IMBH found in M82 plays
the role of ``missing link'' between stellar-mass BHs and SMBHs.

Since we now know that IMBHs exist, it seems natural to expect that
SMBHs might be formed from them. We propose that IMBHs are formed in
the cores of young compact star clusters through merging of massive
stars and BHs formed from them.  These compact young clusters sink to
the galactic center by dynamical friction. At the same time, they
evaporate via thermal relaxation, stellar mass loss and the effect of
the parent galaxy's tidal field. Thus, IMBHs are created and
transported to the center of the galaxy, where they eventually merge
to form SMBHs.

In the following, we discuss how we might seek to confirm our new
scenario. The most direct evidence would be the observation of
gravitational radiation from close binary IMBHs or merging IMBHs. LISA
\citep{Jafryetal1994}, when completed, will be able to detect IMBH
merging events even at cosmological distances. The event rate for
merging of SMBH is estimated to be one per 1--10 years. In our
scenario, each SMBH is a product of $\sim 100$ mergings of IMBHs or IMBH
and growing SMBH. Therefore we predict a much higher event rate for
IMBH-IMBH and IMBH-SMBH merging, of the order of 1 per month or even 1
per week.

To test our hypothesis, searches for IMBHs in other galaxies are
clearly necessary. In our view, IMBHs are likely to form in young
compact star clusters created in nuclear starbursts. We predict that
coordinated observations of nearby starburst galaxies at IR, X-ray and
radio wavelengths, like those performed for M82, will reveal many more
candidate IMBHs.

It is also vital to determine internal and external kinematics of the
host star clusters of IMBHs. High-dispersion spectroscopy in the IR
with large ground-based telescopes such as SUBARU should be able to
determine the velocity dispersion of such a star cluster. Observations
by HST would easily resolve the cluster and give us detailed
information of its structure. Comparison of these results with
theoretical models will then determine whether or not runaway merging
can actually take place there.

The Ultra Luminous Compact X-ray Sources\citep{Mk00,CM99} may be
directly related to IMBHs. These are luminous compact X-ray sources
with Eddington masses exceeding 100 ${\rm M}_{\odot}$.  Most are found
in the arms of spiral galaxies. Thus, they are probably related to
young star clusters, and might well also have been formed through
merging of massive stars.  Their relatively low masses may reflect the
differences sizes and masses of their parent star clusters.

\section*{Acknowledgments}

This work was supported in part by NASA through Hubble Fellowship
grant HF-01112.01-98A awarded by the Space Telescope Science
Institute, which is operated by the Association of Universities for
Research in Astronomy, Inc., for NASA under contract NAS\, 5-26555 and
by the Research for the Future Program of Japan Society for the
Promotion of Science (JSPS-RFTP97P01102).  SLWM and SPZ acknowledge
support of NASA ATP grants NAG5-6964 and NAG5-9264.

\end{document}